\documentclass{aa}
\usepackage[authoryear]{natbib}
\usepackage[a4paper]{hyperref}
\usepackage{amssymb}
\usepackage{graphicx}
\begin{document}

\newcommand{\ms}{$M_{\odot}$}
\newcommand{\msb}{$M_{\odot}$~}
\newcommand{\cd}{$^{12}$C}
\newcommand{\cdb}{$^{12}$C~}
\newcommand{\ct}{$^{13}$C}
\newcommand{\ctb}{$^{13}$C~}

\title{Infrared Photometry and Evolution of Mass-Losing AGB Stars.}

\subtitle{I. Carbon Stars Revisited}

\author{R. Guandalini\inst{1}
\and M. Busso\inst{1}
\and S. Ciprini\inst{2,1}
\and G. Silvestro\inst{3}
\and P. Persi\inst{4} }
\offprints{M. Busso}

\institute{Dipartimento di Fisica, Universit\`a di Perugia, Via
   A. Pascoli, 06123 Perugia, Italy\\
        \email{roald.guandalini@fisica.unipg.it}
     \and
     Tuorla Astronomical Observatory, University of Turku,
     V\"{a}is\"{a}l\"{a}ntie 20, 21500 Piikki\"{o}, Finland \\
    \email{stefano.ciprini@utu.fi}
     \and
      Dipartimento di Fisica Generale, Universit\`a di
        Torino, Via P. Giuria 1, 10125 Torino, Italy\\
        \email{silvestro@ph.unito.it}
     \and
          Istituto di Astrofisica Spaziale e Fisica Cosmica,
          00100 Roma, Italy\\
          \email{persi@rm.iasf.cnr.it}
}

\date{Received 8 April 2005 / Accepted 12 September 2005 }

\authorrunning{Guandalini et al.}

\titlerunning{Infrared Luminosities of C Stars}

\abstract{ As part of a reanalysis of galactic Asymptotic Giant
Branch (AGB) stars at infrared (IR) wavelengths, we discuss a
sample (357) of carbon stars for which mass loss rates, near-IR
photometry and distance estimates exist. For 252 sources we
collected mid-IR fluxes from the MSX (6C) and the ISO-SWS
catalogues. Most stars have spectral energy distributions up to
21~$\mu$m, and some (1/3) up to 45~$\mu$m. This wide wavelength
coverage allows us to obtain reliable bolometric magnitudes. The
properties of our sample are discussed with emphasis on $\sim$ 70
stars with astrometric distances. We show that mid-IR fluxes are
crucial to estimate the magnitude of stars with dusty envelopes.
We construct HR diagrams and show that the luminosities agree
fairly well with model predictions based on the Schwarzschild's
criterion, contrary to what is widely argued in the literature. A
problem with the brightness of C stars does not appear to exist.
From the relative number of Mira and Semiregular C-variables, we
argue that the switch between these classes is unlikely to be
connected to thermal pulses. The relevance of the two populations
varies with the evolution, with Miras dominating the final stages.
We also analyze mass loss rates, which increase for increasing
luminosity, but with a spread that probably results from a
dependence on a number of parameters (like e.g. different stellar
masses and different mechanisms powering stellar winds). Instead,
mass loss rates are well monitored by IR colours, especially if
extended to 20~$\mu$m and beyond, where AGB envelopes behave like
black bodies. From these colours the evolutionary status of
various classes of C stars is discussed.

}
\maketitle

\section{Introduction}

Winds from AGB stars replenish the Interstellar Medium with a
large portion of the matter returned from stellar evolution
\citep[70\% according to][]{sedlmayr94}, through the formation of
cool envelopes \citep{winters02, winters03} where dust grains
condense \citep{schirr03, carciofi04}. Such grains carry the
signature of the nucleosynthesis episodes occurring in the AGB
phases \citep{busso99, was05}; their presence has been recognized
in meteorites and brings direct information on circumstellar
processes \citep[e.g.][]{zinner00,ott01}. As AGB stars radiate
most of their flux at long wavelengths, large surveys of infrared
(IR) observations play a fundamental role in studying their
luminosity and their winds \citep[see e.g.][]{habing96,epch1}.

\begin{figure}[t!]
{\includegraphics[width=\columnwidth]{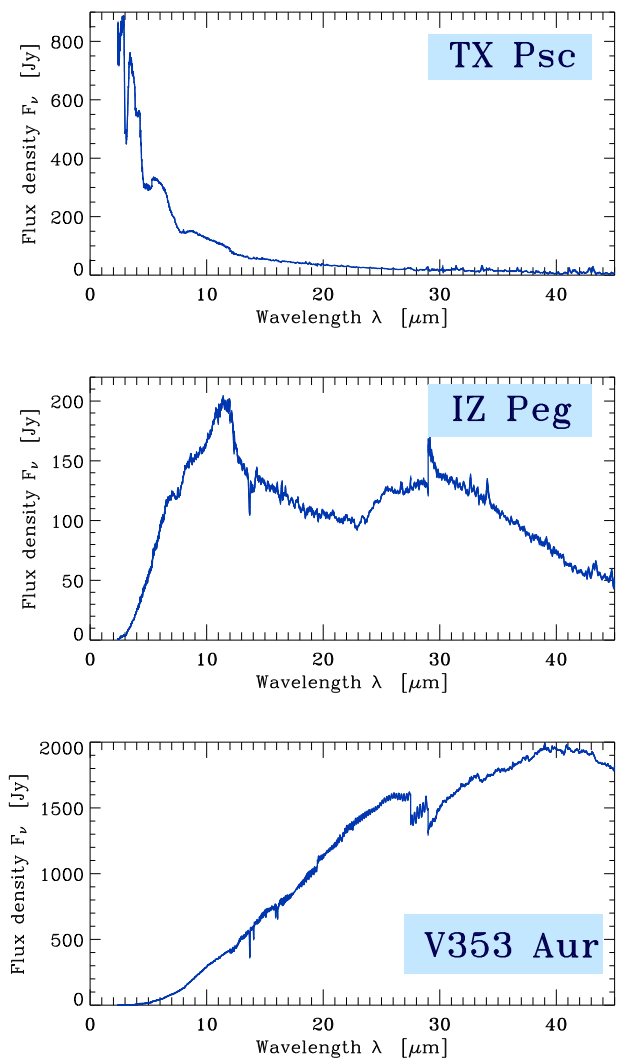}}
    \caption{SWS spectra for an irregular variable (TX Psc),
    a Mira variable (IZ Peg)
    and a post-AGB star (V353 Aur). The dominant role of IR
    emission longward of 20~$\mu$m for Miras and post-AGB sources
    is a general property of our sample.}
              \label{Fig1}%
    \end{figure}

We do not have yet a quantitative description of AGB winds, though
attempts at modelling them are a few decades old
\citep{salpeter,knappmorris,gail}. More recently, hydrodynamical
and phenomenological studies of pulsating stellar atmospheres and
of the associated mass loss have undergone important improvements
\citep{fleischer92,woodsebo,winters02,winters03,wachter02,sandin3a,sandin3b}.
Also, observational works have become more quantitative
\citep{wood03,olivier,wood04,andersen03}, using new data at long
wavelengths from space and from the Earth \citep[see
e.g.][]{woodcohen,lebertre01,lebertre03,groenewegen02a,groenewegen02b,olofsson03,cioni03,omont03},
as well as improved knowledge of stellar distances
\citep[e.g.][]{vaneck,knapp03,bergeat05}.

A real step forward would be to derive realistic formulae, linking
the efficiency of stellar winds to the luminosities, colours and
chemical properties of AGB stars, to be adopted as inputs for
stellar models, thus avoiding free parameterizations. This kind of
studies has become quantitative only recently \citep{vanloon05},
while previous attempts \citep{vassiliadis,blocker95} suffered
from large uncertainties \citep{wood96}.

Reducing those previous uncertainties also requires a good
knowledge of absolute magnitudes, through improved distances; so
far, mass loss studies for galactic AGB stars often adopted some
{\it average} constant value for the luminosity
\citep[see][]{jura86,jura89,lebertre01}. Relevant exceptions
exist, e.g. in the thorough analysis of C-rich sources by Bergeat
and his group \citep[see][ hereafter BKR1, BKR2]{bergeat02a,
bergeat02b}, where however the photometric data are rather
heterogeneous \citep[see also][]{knapik99,bergeat01}.

With the above difficulties in mind, we plan to perform a thorough
reanalysis of AGB luminosities and mass loss: this is actually the
main scientific scope of our project for putting an IR telescope
(the International Robotic Antarctic Infrared Telescope) in
Antarctica, at the Italian-French base of Dome C \citep{tosti}. It
will allow extensive monitoring of AGB sources in Magellanic
Clouds and in the southern Milky Way up to 20~$\mu$m (and possibly
beyond, given the unique characteristics of the Antarctic
atmosphere). This paper is in fact part of the preliminary works
necessary to define the key projects for the telescope.

We shall therefore consider, in a series of works, all types of
AGB stars (M, MS-S, C giants), using existing catalogues of IR
observations and compiling a homogeneous list of luminosities,
distances, mass loss rates, looking for correlations between them
and the IR data. We start with C-rich AGB stars; next steps will
analyze MS-S giants, where C- and $s$-element enrichment becomes
observable, and finally the more disperse family of M giants.

In this note we collected a sample of 357 C-rich sources with
near-infrared (IR) photometry and mass loss estimates; these last
are revised by scaling previous observations with updated distance
determinations. For another sample of 252 C stars we have
collected also space-borne mid-IR photometry in the $\lambda =
8-15 \mu$m interval and/or at $\lambda \sim 21 \mu$m. The
intersection of these two groups is made of 214 C-rich objects.
For this sample, near IR data are provided by ground-based surveys
like 2MASS \citep{cutri} and DENIS \citep{epch0,fouque}; those for
mid-IR are taken from the recent space-borne missions, ISO and
MSX. The relevance of MSX colours for classifying AGB stars has
been recently pointed out \citep{ortiz05}, with an approach which
is complementary to ours. IRAS measurements are not included in
our analysis because of their lower quality and spatial resolution
(hence larger contamination) compared to more modern experiments.
A homogeneous compilation of IR colours for stars of known
distance will allow us to precisely estimate the bolometric
corrections (BCs) and to correlate mass loss rates with
luminosity (and/or the colour excess).

We started from previous work done by BKR1 \& BKR2, who classified
C stars in several discrete families: the more advanced
subclasses, named 'Carbon Variables' or CVs, from CV3 on,
correspond to bright thermally pulsing AGB stars. Bolometric
magnitudes were estimated by those authors using optical
photometry by \citet{baumert}, near-IR data from \citet{gezari},
and including IRAS fluxes at 12 and 25~$\mu$m. Careful upgrades of
Hipparcos were included. In our work we shall consider newer and
more homogeneous IR data from the ground and from space, with
Spectral Energy Distributions (SEDs) covering a wider wavelength
baseline up to 45~$\mu$m. In this way we also aim at verifying on
wider statistics suggestions previously advanced by
\citet{busso96}, \citet{marengo97,marengo99}, \citet{busso01},
according to which mid-IR colours are good indicators of the mass
loss efficiency, and permit a first classification of the chemical
properties of the circumstellar envelopes.

The choice of discussing first a sample of Carbon-rich AGB stars
is motivated by the fact that optical C(N) stars (usually
irregular or semi-regular pulsators) present a sufficient level of
homogeneity to be considered as a snapshot on the relatively long
(1-3$\times$10$^6$ yr) TP-AGB evolution. They correspond to the
moment in which the C/O number ratio reaches unity in stars of
moderate mass \citep[lower than 3
\ms:][]{claussen,busso99,abia1,abia2,kahane}. When looking at IR
properties, however, one has to remember that more massive and
evolved sources become part of the sample, thus making the picture
more complex \citep{barnbaum}. Mass loss rates have been shown to
span a wide interval, from a few 10$^{-8}$ \ms/yr,
\citep{olofsson93a,olofsson93b,schoier01} to 10$^{-4}$ \ms/yr for
the most massive and/or evolved objects
\citep{groenewegen02a,groenewegen02b}.

All stars outside the mass range in which C-rich atmospheres are
formed will reach the final stages while remaining O-rich. This is
due either to the inefficiency of dredge-up (for lower masses), or
to the fact that massive AGB stars ($M$ $\ge$ 5 \ms) burn the new
carbon as soon as it is dredged to the envelope, thanks to the hot
bottom burning (hbb) phenomenon \citep{karakas4}. Observational
evidence of this occurrence is emerging in various contexts
\citep{smith,vanloon01,vanloon05,whitelock,cioni03}. See in
particular \citet{vanloon99a,vanloon99b}, hereafter VL1, VL2. This
paper is organized as follows. In section 2 the C-star sample and
the photometric measurements are presented; in section 3 we
discuss mass loss rates from radio studies, together with the
required updates (e.g. on the distance). The luminosities and
colours of our sample stars are presented in sections 4 and 5,
together with the consequences of our analysis for the AGB
evolution and dredge up. Correlations found between luminosity,
colours and mass loss rates are reviewed in sections 6. Some
general conclusions are drawn in section 7.

\section{C-star photometry from ISO-SWS and MSX catalogues}

Ground-based photometric studies of C stars in the 8-14 ~$\mu$m
window were often performed through 1$\mu$m-wide filters, centered
around 8.8, 11.7 and 12.5 ~$\mu$m. We decided therefore to compile
colours including filters as close as possible to the above ones,
to make comparisons easier.

The largest inventories of IR observations longward of 2.2 ~$\mu$m
for AGB stars come from space experiments.
\citet{lebertre01,lebertre03} examined the sample from the
Japanese experiment IRTS. We shall consider instead data from the
European ISO mission and from the American MSX satellite (using
the Infrared Point Source Catalogue, version 6C).  In the case of
ISO, the best known, nearby AGB stars are always too bright to be
observed by the imaging camera \citep[ISOCAM: see][]{cesarsky}.
Luckily, the database of the Short Wavelength Spectrometer
\citep[SWS:][]{valentijn} contains spectra of many known AGBs, and
offers also a good photometric calibration. We therefore performed
a re-binning of AGB SWS spectra from the ISO archives at the {\it
Centre de Donn\'ees Stellaires (CDS)} of Strasbourg. In doing
this, the spectral energy distributions were convolved with the
response curves of common filters centered at 8.8, 11.7 and 12.5
~$\mu$m, available from the TIRCAM consortium \citep{persi}. For
the sake of simplicity, these filters will be indicated by: [8.8],
[11.7], [12.5]. We also rebinned data at longer wavelength,
convolving them with the filters at 14.7 and 21.3 ~$\mu$m from
MSX. Sources for which ISO fluxes in different epochs vary
significantly were combined to get average spectra: a few sources
showed the remarkable property of a global (bolometric) variation
up to a factor of two. Such phenomena have been observed
previously \citep{lebertre92,lebertre93,vanloon98}: for
interpretations the reader is referred to \citet{vanloon98}.

A number of our sources have complete SWS spectra up to 45~
$\mu$m: those among them that also have near-IR photometry and
estimates for mass loss and distance (49 in total of the about 70
astrometric sources) represent a 'best selection', from which we
can expect to derive all photometric properties with high
precision, including the bolometric magnitudes (these last will
simply come out from integrating the flux from 1 to 45~$\mu$m, as
for the whole sample optical fluxes are  very small compared to
the IR ones). Three examples of such complete spectra,
representative of stars with small, moderate and large IR excess,
are presented in Fig. \ref{Fig1}. They show one of the main points
we want to underline in this work: deriving bolometric magnitudes
from too short wavelengths can be very misleading, especially for
very red sources. As examples, using BCs derived from J--K, as in
\citet{costa}, the sources in Fig. \ref{Fig1} would be assigned
absolute magnitudes of: $-4.8$ (TX Psc); $\ge 1.0$ (IZ Peg); $\ge
1.0$ (V353 Aur). The last two numbers derive from assuming
generous upper limits of 0.05 Jy for the very low J fluxes of
these red sources. When BCs are instead computed starting from IR
photometry extending from the K-band to 10--20~ $\mu$m, as in
\citet{lebertre01}, more reasonable magnitudes are obtained:
$M_{bol} = -5.2, -4.8, -5.9$, respectively. When, finally, we
apply our BCs (see next sections) derived from photometry up to
45~ $\mu$m, we get $M_{bol} = -5.5, -4.8, -6.2$ in the three
cases. For very red sources (about 25\% of Miras and most
post-AGBs) even our Spectral Energy Distributions (SEDs) up to 45~
$\mu$m are insufficient to gather the whole flux. For them, the
luminosities we estimate will be lower limits.

Concerning the MSX sample, we adopted the fluxes at 14.7 and 21.3
$\mu$m as provided by the catalogue, while in the 8--14 ~$\mu$m
range, fluxes in the [8.8] and [12.5] filters were inferred from
those in the MSX bands 'A' and 'C'. In order to do this we
selected a group of sources in common between ISO-SWS and MSX. For
them, we derived correlations between the two sets of filters,
obtaining the following relations:
$$
\log F_{12.5} = 0.80176 ~\log F_C + 0.3349
$$

$$
\log F_{8.8} = 0.9377~\log F_A + 0.1266
$$
with regression coefficients ($R^2$) of 0.95, 0.94 respectively.
The relations apply to all sources examined: colour effects on
different types of stars are negligible. They were used for
expressing the MSX fluxes in the set of filters adopted. This can
be done without loss of accuracy, as the uncertainty of IR colours
is at least $\pm$ 0.1$^m$. No flux at 11.7 ~$\mu$m can obviously
be derived for MSX stars.

The resulting list of ISO and MSX C-rich AGB sources, with their
inferred IR fluxes, is presented in Tables \ref{tab1} to
\ref{tab4}. Here variability types are taken from the Combined
General Catalogue of Variable Stars \citep{samus}. In the tables
we have included near-IR data from 2MASS, together with general
information available on the variability type and on parameters
relevant for estimating mass loss rates (see next section). One
problem affecting our sample is that data from catalogues are
single-epoch measurements; though the amplitude of the light
curves decreases sharply when moving from optical to near-IR
wavelengths, we expect that J, H, K data can be still affected by
uncertainties of several tenths of a magnitude. This problem
affects only the bluest semi-regular and irregular variables (at
the lowest luminosities and mass loss rates considered). Indeed,
despite the fact that redder sources also show larger amplitude
variations in near-IR filters, their flux is usually dominated by
mid-IR wavelengths, where the variability vanishes because the
photosphere is not seen. The very large near-to-mid IR colours of
several Miras and of all post-AGB sources make any variability at
short wavelengths irrelevant (exceptions might occur for the few
sources showing factor-of-two variations in the bolometric
luminosities). Also, the photometric data in the tables do not
contain corrections for interstellar extinction, for two reasons.
On one side extinction becomes small at the long wavelengths
considered here; on the other hand, our stars are distributed in
all directions of the sky, so that we should adopt average
extinction laws, but these are rather uncertain, and it is not
clear that (at IR wavelengths) average corrections are
substantially better than nothing. As our stars are occasionally
very distant (above 1 kpc) one has to estimate the effects of
neglecting extinction. Let's consider representative cases at 200
and 2000 pc distance. For the average interstellar medium one has
$A_V$(200pc) = 0.15$^m$; $A_V$(2000pc) = 2.2$^m$
\citep{gillett,knude}. The wavelength dependence can be taken from
\citet{draine}, as being $A_{\lambda} \sim {\lambda}^{-1.75}$. At
mid-IR wavelengths (10~$\mu$m and beyond) this implies negligible
corrections (0.015$^m$ at maximum), well within the photometric
uncertainly, even for the most distant sources. In the K-band (2.2
~$\mu$m) extinction is irrelevant within 1kpc, but may be
appreciable for the most distant stars (0.2$^m$ at 2 kpc). We
notice that the unknown variability induces a scatter whose
amplitude is similar to the uncertainty related to extinction.
Neglecting this last, however, introduces systematic effects on
the most distant stars, so that their near-IR luminosity will be
underestimated. Hence at near-IR wavelengths, catalogue data for
distant ($\geq$ 1 kpc) LPVs taken at face values are typically
uncertain by a few tenths of a magnitude (from both statistical
and systematic errors). The effects on the bolometric magnitude
distribution in our sample are however small, due to the large IR
excesses of most distant sources that make near-IR data relatively
unimportant for them.

When using the data of Tables \ref{tab1} to \ref{tab4} for
determining standard colours, we applied standard photometric
calibrations, adopting suggestions by \citet{bessell} and by
\citet{glass} and interpolating them with black bodies for
obtaining the fluxes in the selected 1-$\mu$m-wide filters.
According to this procedure, the fluxes (Jy) of a zero-magnitude
star in the filters used are: 52.23 ([8.8]), 29.55 ([11.7]), 25.88
([12.5]), 20.25 ([14.7]), 8.91 ([21.3]). For near-IR, 2MASS
calibrations are given in \citet{cohen}.

\section{Estimates for mass loss and related problems}

The work by \citet{jura86} and \citet{jura89} set the stage for
subsequent studies on AGB winds. These authors used mass loss
rates estimated across the previous ten years, and mid-IR
photometric observations from the IRAS satellite. Unfortunately,
the scarcity of data then available led them to adopt {\it
average} hypotheses on the luminosities, assuming a constant value
of 10$^4$ $L_\odot$ for all types of AGB stars. This fact (and the
modest precision of IRAS photometry) prevents us from making too
systematic comparisons of those works with the present analysis,
where the luminosities are estimated thanks to measurements of the
distance, sometimes of astrometric quality. We shall however
comment later on mass loss rates for the sources we have in common
with the quoted studies.

The inventory of more recent analyses on mass loss in AGB stars
includes works dedicated to Carbon-rich stars
\citep{olofsson93a,olofsson93b,kastner93,schoier01,groenewegen02a,groenewegen02b,schoier02},
to Oxygen-rich M--MS--S stars
\citep{sahai95,groenewegen98b,olofsson02}, and more general
approaches including all classes of AGB objects
\citep{kastner92,loup,lebertre01,lebertre03,winters02,winters03}.
From the theoretical point of view, the very limited evidence of
coronal X-ray emission \citep{jura84,sahai03,kastner03}
demonstrated that mass loss mechanisms were different than in the
sun. Radiation pressure on dust grains was suggested to be at the
origin of red giant winds by \citet{salpeter}. In this approach,
if the momentum of the radiation field can be completely
transferred to dust and gas particles in the envelope, then the
mass loss rate can be expressed as:
$$
\dot M \sim 2\times 10^{-8} \tau_d (L_*/L_{\odot})\times(v/km
s^{-1})^{-1} M_{\odot}/yr
$$
where $\tau_d$ is the average optical depth of dust. The gas phase
of the envelope might be dragged by dust, but a problem is how to
transfer momentum to dust close to the photosphere, as above
1500-2000 K no solid compound made with light elements (C to Si)
would avoid evaporation. Hence, stellar pulsation in different
modes \citep{wood99,wood00}, pushing the radius to large distances
and the temperature to low values, and introducing shock waves
should also be very important \citep[see
e.g.][]{fleischer92,fleischer95,winters94a,winters94b,winters97,winters00a,winters00b}.

All approaches suffer from many limitations, so that recent work
was concentrated on semi-empirical approaches derived from
observations. As an example, \citet{lebertre01,lebertre03} used
near-IR fluxes (in K, L'), then adopting simplifying assumptions
like e.g a constant luminosity of 8000 $L_{\odot}$ for all the
sources studied. Being based on a wavelength interval where the
AGB photosphere dominates the flux, these estimates are also
affected by the intrinsic uncertainty of the variability.

Other mass loss estimates were derived from observations at
millimeter or radio wavelengths, where the variable star is not
seen. In most cases they adopt procedures early suggested by
\citet{knappmorris} for CO lines. For example,
\citet{olofsson93a,olofsson93b} and
\citet{groenewegen02a,groenewegen02b} assume emission from an
optically thick environment, with a fixed size for the CO
envelope, $R_{CO}$. Due to this, they can sometimes underestimate
the wind efficiency by factors 3-4 \citep{schoier01}.

A more general case was considered by \citet{loup},
\citet{kastner92}, \citet{kastner93} and \citet{winters03}. The
results are given in terms of the distance $d$ (in kpc), of the
expansion velocity $V_e$ (in km/sec), and of the flux $f$. We
adopt \citet{loup}'s formulation whenever possible, re-scaling
their mass loss rates with more recent estimates for the distance
and for $V_e$. When data from \citet{loup} are not available, we
use instead the \citet{olofsson93a,olofsson93b} procedure, again
updating the parameters when possible.

\begin{figure}[t!]
   \centering
   {\includegraphics[width=\hsize]{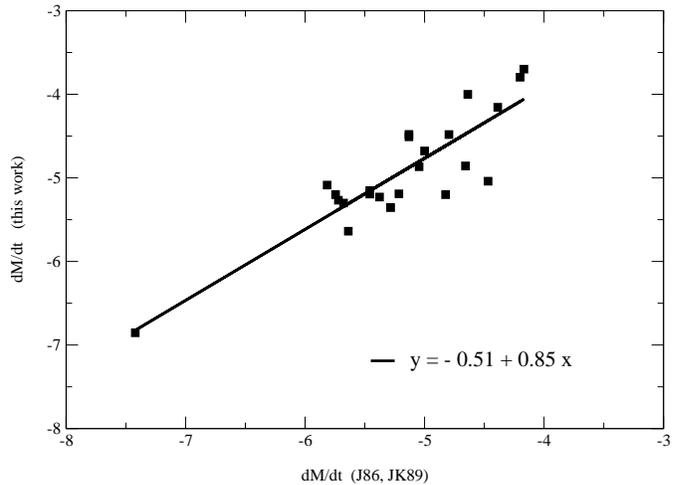}}
    \caption{The existing correlation between mass loss estimates published by \citet{jura86}
    and \citet{jura89} and our re-calibration of measurements from
    \citet{loup} with updated distances.}
    \label{Fig2}%
    \end{figure}

Tables \ref{tab1} to \ref{tab4} report the relevant data for those
stars for which IR photometry was derived in the previous
section, together with the updated estimates for mass loss rates.
It is interesting to note (see Fig. \ref{Fig2}) that for the few
sources we have in common with \citet{jura86} and \citet{jura89}
and despite the mentioned uncertainties of these older analyses,
there is a fair correlation between the mass loss rates given by
the above authors and those derived by us (with a regression
coefficient $R^2$ = 0.88). In most cases the sources are Mira
variables.

We shall be guided by the data of the AGB stars (Table \ref{tab1})
with the best distance estimates. They include objects with
astrometric measurements (usually from Hipparcos), for which we
adopt the recommended values by \citet{bergeat05}, plus a small
group of post-AGB stars for which the distance was derived from
dedicated studies of the circumstellar envelopes. Our mentioned
'best selection' is actually made by those sources, among the
objects in Table \ref{tab1}, for which also the integral of the
SED up to 45~ $\mu$m is available. We give a lower weight to other
methods of deriving distances (the corresponding points in the
figures will be smaller). When nothing else is available, we quote
in the tables distances estimated by \citet{loup} by assuming a
specific value for the luminosity. Obviously, these sources will
be used only when a precise knowledge of the distance is
irrelevant, e.g. in colour-colour diagrams.

\section{Energy distributions and photometric calibrations}

\begin{figure}[t!]
   \centering
   {\includegraphics[width=\hsize]{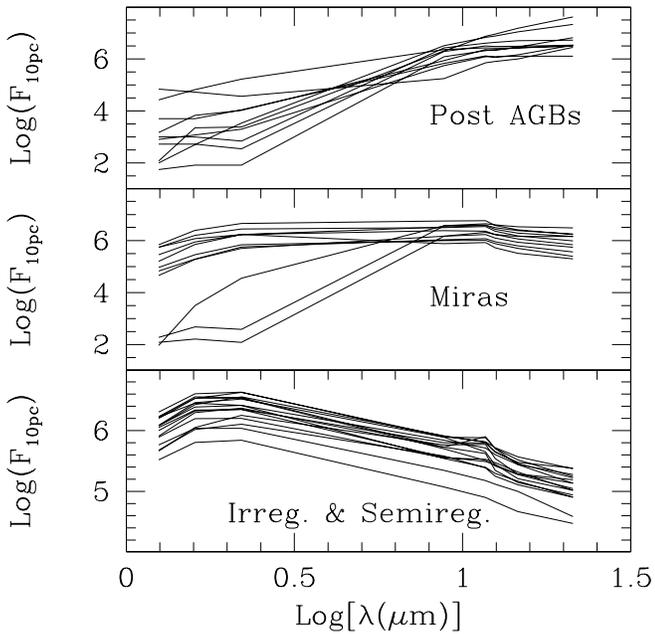}}
\caption{The Spectral energy distribution up to 21.3~ $\mu$m of a
number of sources in our sample, divided according to their type,
as constructed with the set of 1-$\mu$m-wide filters described in
the text.}
\label{Fig3}%
    \end{figure}

A few SEDs, chosen to represent the various shapes found in our
sample, are shown in Fig. \ref{Fig3}. They illustrate how
different the spectra for different types of sources are,
displaying the increasing importance of mid-IR (10 and 20~ $\mu$m)
wavelengths when moving from irregular (or semi-regular) pulsators
toward Miras and post-AGB stars. This has two important
consequences as discussed below.

First of all, it is clear from the figure that, when the
observations are not extended up to at least 20 ~$\mu$m, most of
the flux from the reddest objects is missed. Plots like those in
Fig. \ref{Fig1} reveal that many AGB stars radiate the majority of
their flux in a region between 20 and 45 ~$\mu$m. This range is
rarely considered in dealing with large samples of sources. In the
past, the flux in the 60~ $\mu$m IRAS filter was often used as
representative of far IR, and correlated with mass loss. However,
IRAS 60~$\mu$m data may be partially contaminated by diffuse
(cirrus) emission; moreover, the IRAS spatial resolution at those
wavelengths is very poor.

\begin{figure}[t!]
   \centering
   {\includegraphics[width=\hsize]{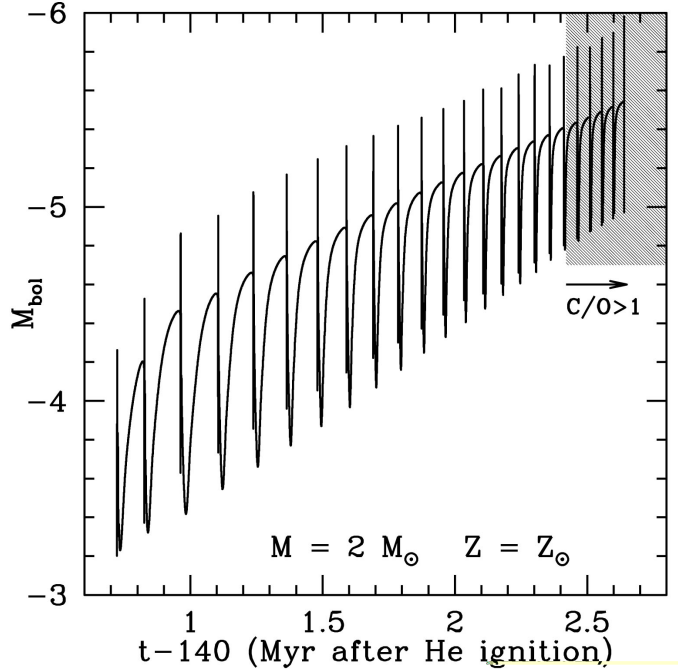}}
\caption{Model bolometric magnitude of a 2 $M_{\odot}$ AGB star
during the phase of thermal pulses on the AGB. Model from the
FRANEC code, as computed by \citet{straniero98}.}
\label{Fig4}%
    \end{figure}

Secondly, when we consider the relative number of sources having
SEDs with the various appearances of Fig. \ref{Fig3} (with
semiregulars and Miras occurring with nearly equal frequency)
doubts arise on the common idea that the different variability
types are alternatively encountered along the AGB due to the
variations of luminosity in correspondence of thermal pulses
\citep[see e.g.][]{cioni01}. These doubts are best illustrated
with reference to the behaviour of the stellar luminosity on the
AGB (Fig. \ref{Fig4}). The usual interpretation might in principle
be true for the first thermal pulses, when the star is
oxygen-rich, as regimes of higher and lower luminosity are both
encountered, due to the occurrence of post-flash dips. In early
phases these dips cover nearly 30\% of the time between adjacent
pulses \citep[for a recent review on the models, see
e.g.][]{straniero05} and hence imply that lower-luminosity
variables (identified with semiregulars) should account for $\sim$
30\% of the total number of O-rich AGB stars. However, for the
final C-rich phases the duration of low-luminosity regimes becomes
increasingly short with increasing pulse number. According to Fig.
\ref{Fig4}, relative to a 2 $M_{\odot}$ AGB star of solar
metallicity, the stages where the luminosity is lower than average
by at least 0.2$^m$ account for about 10 percent of the time spent
by the star as a C-rich giant: if C-rich semiregular variables
were to be associated with these low luminosity regimes, they
should be pretty rare, contrary to the evidence we have. Our
sample was selected from mass loss data and therefore is certainly
biased toward the reddest and more mass-losing Miras, but despite
this semiregular variables are of equal statistical weight. If the
AGB stars really switch from one variability type to another
\citep[see e.g.][]{cioni01} this is unlikely to be related to
thermal pulses; rather, in atmospheres animated by complex
pulsations, superposition of close-by frequencies can induce beats
and hence amplitude modulations \citep[with periods of centuries
or millenia, see e.g.][]{marengo01} that may be independent from
what happens in the internal layers. One has to notice that, in
the BKR1 \& BKR2 classification in CV classes, the complex
behaviour of the C/O ratio shows an increase with class number and
with decreasing $T_{eff}$, down to about 2500K; Miras (classes
CV5$-$CV6) occupy regions with higher C/O ratios than SRa,b stars
(classes CV3+ and CV4). This fact, and the gradual SED changes we
see (Fig. \ref{Fig3}), with smooth transitions from Semiregulars
to Miras and then to post-AGBs, leads to the conclusion that, at
least on average, most Miras are in an evolutionary stage
subsequent to semiregulars. Oscillations between the two types
should exist, but the final evolutionary stages should be more
heavily populated by Mira-like variables.

SEDs like those of Figure \ref{Fig3} were the base for estimating
bolometric magnitudes. We used for this the fundamental relation
of photometry, according to which one has \citep[see
e.g.][]{glass}:
\begin{equation}\label{1}
  M_{bol} = \int_{0}^{\infty}{F_{\nu}d\nu} + C
\end{equation}
where C = $-$18.98 when the flux is expressed in W/m$^2$. As a
consistency check we derived the zero-magnitude flux from the data
of \citet{bessell}, extended from the U to the Q band, finding
C=$-$19.01, a very good approximation to (1). The bolometric
magnitude of AGB stars has often been derived from corrections
applied to the K magnitude; these corrections are usually a
function of the K--[12] colour \citep[using e.g. IRAS data,
see][]{lebertre01,lebertre03}. For the sake of comparison, we too
present here our $M_{bol}$ estimates through bolometric
corrections as a function of the (K--[12.5]) colour. If $f$ is a
measure of the flux in Jy (Tables \ref{tab1}--\ref{tab4}):
\begin{equation}\label{2}
  K = -2.5\times \log~f_{\nu,K} + 5 - 5\times \log~d
  +2.5\times \log~f_{\nu,K_{0}}
\end{equation}
where $d$ is the distance in parsec and $f_{\nu,K_{0}}$ is the
zero-magnitude flux in K.

\begin{figure}[t!]
   \centering
   {\includegraphics[width=\hsize]{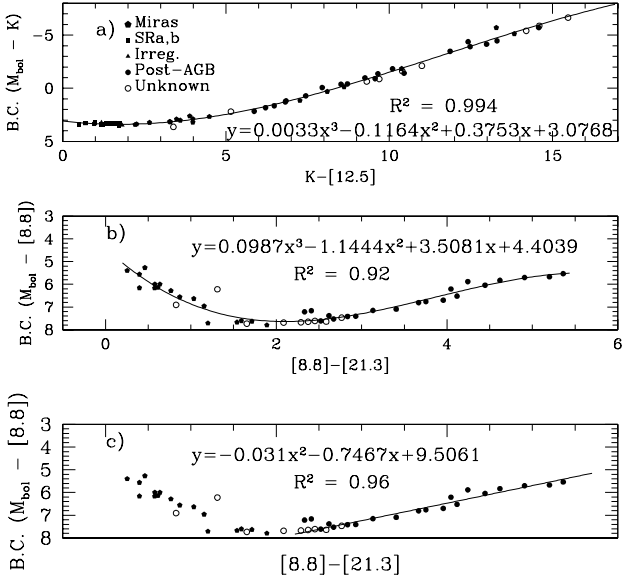}
    }
    \caption{Panel a): Bolometric corrections for the K magnitude as a function
    of the K--[12.5] colour. They were derived
    for AGB C-stars with complete SEDs, from 2MASS and ISO-SWS, up to 45~
    $\mu$m. Panels b) and c): bolometric corrections for the [8.8] magnitude, as
    a function of the mid-IR colour [8.8]--[21.3]. Of the two
    relations shown, the one in panel c) is more appropriate for
    extrapolations to very red sources, as the cubic spline of panel b)
    introduces unwanted biases outside the range shown.}
\label{Fig5}%
    \end{figure}

\begin{figure}[t!]
   \centering
   {\includegraphics[width=\hsize]{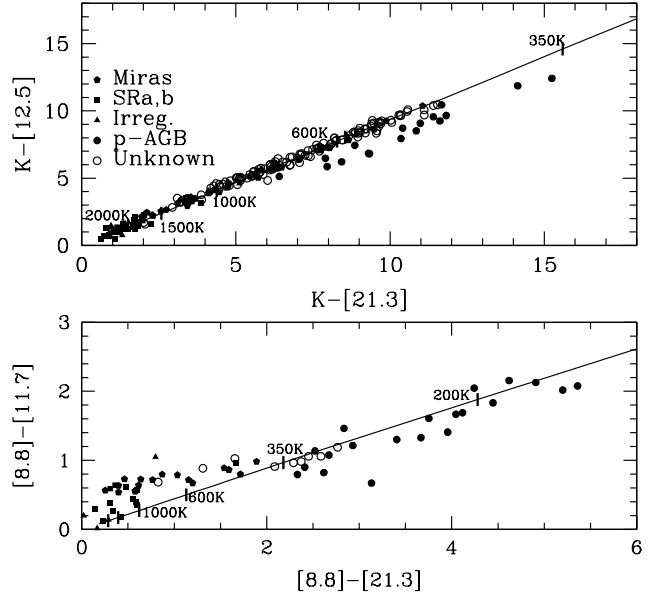}}
    \caption{Examples of colour-colour diagrams in the IR for C-rich AGB
    and post AGB stars.}
\label{Fig6}%
    \end{figure}

Bolometric corrections derived from equations (1) and (2) are
shown in Fig. \ref{Fig5} (panel a) as a function of the K--[12.5]
colour, together with a fitting formula for extrapolations.
Optically-selected sources (in general semi-regular variables)
have small corrections: their flux is well estimated from
traditional criteria at short wavelengths, as most flux is
radiated in near-IR. Alternative expressions of the B.C., as a
function of mid-IR colours alone, to be used for very red sources
too faint in near-IR, are presented in Fig. \ref{Fig5} (panels b,
c). Here semiregular variables are omitted, because of their small
IR excess (see Fig. \ref{Fig1}).

Finally, we must notice that mid-IR emission from circumstellar
envelopes makes the stars of our sample appear as black bodies at
long wavelengths, apart from the known typical features of C-rich
dust. This is shown in Fig. \ref{Fig6}, where calibrated
colour-colour diagrams are shown for near- and mid-IR filters. In
the top panel, adopting the K$-$[21.3] colour as baseline, the
data points of known AGBs and of sources (mainly from MSX) of
unknown variability type are almost perfectly aligned on the
blackbody sequence; they should be dominated by cool, evolved
Mira stars with extended dusty envelopes; post-AGB sources
instead display excess emission at 21.3~ $\mu$m, which is a
well-known property of such C-rich objects \citep{vanwinckel}.
The bottom panel then shows that, if only mid-IR is used,
including the [11.7] filter where SiC and PAH features are
present, then normal AGB C stars show excesses in this filter,
while post-AGB C-rich sources gradually move to an excess at 21.3
$\mu$m, while becoming redder. In general, Miras are redder than
semiregulars and Post-AGBs are the reddest sources of the sample
(due to their cold dust). In the evolutionary hypothesis we have
tentatively advanced before (at least on average), this graph
would offer an immediate tool for classifying the evolutionary
status of C-rich circumstellar envelopes.

\section{Bolometric magnitudes, colours and model comparisons}

\begin{figure}[t!]
   \centering
   {\includegraphics[width=\hsize]{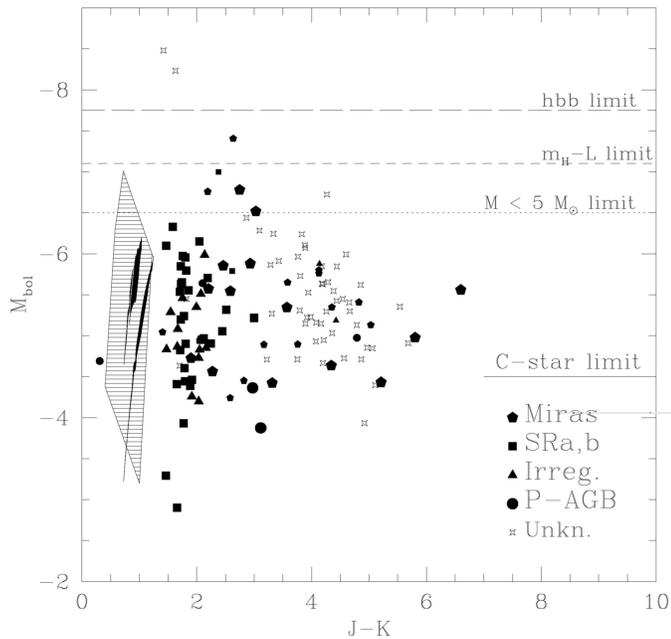}}
\caption{The HR diagram of observed C stars in near IR, as
compared to the area (dashed) covered by canonical stellar models
(without hbb). The minimum limit for C star occurrence in the
adopted models is indicated}
\label{Fig7}%
    \end{figure}

\begin{figure}[t!]
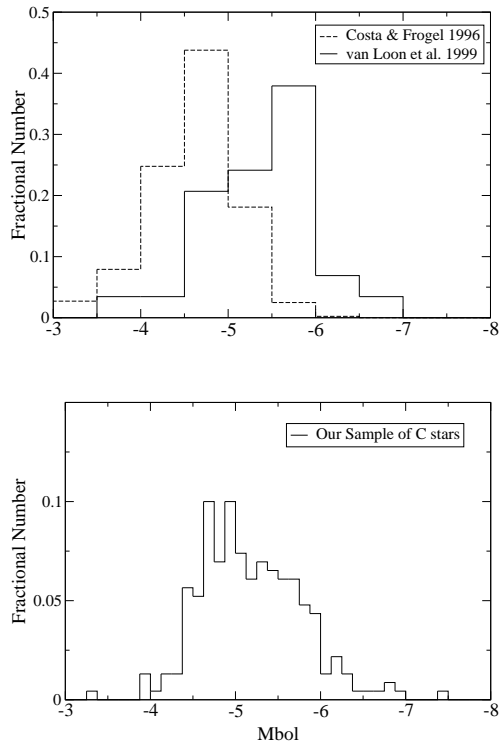

   \centering
    {\includegraphics[width=6.5cm]{3208f08a.eps}}

\vspace{0.5cm}

{\includegraphics[width=6.5cm]{3208f08b.eps}}
    \caption{Top panel: the histogram showing the fractional number of C stars
    per magnitude interval in our sample (230 sources). Bottom panel: same as before,
    for the samples of LMC C stars in optical wavelengths \citep[][ 887 sources]{costa}
    and in the infrared (VL2, 29 sources; note that the bin
    size here is the same as used by these authors for their data.}
\label{Fig8}%
    \end{figure}

The magnitudes derived by our analysis are used in Fig.
\ref{Fig7}, where HR diagrams are presented, using an IR colour as
abscissa. As stated, small symbols refer to stars with
less-accurate distances: they are usually distant post-AGB
objects, or AGBs of unknown variability type. The dashed area in
Fig. \ref{Fig7} represents the zone covered by standard stellar
models with the Schwarzschild's criterion for convection, adopting
the usual mixing length theory as adapted to stellar physics by
\citet{cox}. The value of the so-called parameter $\alpha$ (ratio
of the mixing length to the pressure scale height) is $\alpha$ =
2.1 \citep{straniero98}. All models considered are from the FRANEC
evolutionary code; they refer to metallicities from
Z=Z$_{\odot}$/3 to Z=Z$_{\odot}$ and include models of relatively
low mass (2 to 3 \ms) \citep{straniero7, gallino98, busso99}, plus
intermediate mass star models \citep{dominguez, straniero00}. As
representative cases, we plot explicitly the track of a 2\msb star
with Z=Z$_{\odot}$, and that of a 3 \msb star with Z=Z$_{\odot}$/3
(heavy lines). As compared to the evolutionary tracks, the data
points (which are systematically displaced at redder colours)
show the gradual formation of cool and dusty circumstellar
environments, which recycle part of the stellar flux at longer
wavelengths through the absorption and re-emission mechanisms of
dust grains. For the bluest sources, the displacement in J--K
with respect to theoretical predictions might also be due to
models underestimating molecular opacities of C-rich material
\citep{mar03}.

We underline that the models adopted do not assume any overshoot
for convection \citep[contrary to, e.g.][]{frost6}; for the
techniques used to compute the convective border see
\citet{stran3}. Overshooting from the convective envelope and in
general an extension of dredge up larger than allowed by the
Schwarzschild's criterion is even today a common assumption
\citep{herwig05}. This is usually claimed to be needed for letting
the photospheres become C-rich {\it at low luminosities}, in order
to fulfill constraints like those by \citet{blanco} and
\citet{costa}, derived by optical and near-IR data, suggesting
average C-star magnitudes below $-$5. We notice that similar works
in the optical bands became recently feasible on other galaxies,
and the results were essentially the same, yielding C-star
magnitudes between $-$4.5 and $-$5, again neglecting IR fluxes
\citep{bat04a,bat04b,demers}.

Despite some contrary evidence, which has emerged from IR studies
of Magellanic Clouds and of the Galaxy (see \citet{vanloon98},
VL1, VL2, BKR2) some theorists assume, even today, that such faint
bolometric magnitudes characterize the whole family of C-stars,
and yield a disagreement with stellar models adopting the
Schwarzschild's criterion \citep{stancliffe,herwig05}. Our data do
not support this belief: Fig. \ref{Fig7} reveals a substantial
agreement of the observed luminosities with model predictions.
Indeed, observed bolometric magnitudes are mainly concentrated in
the range from $-$4.5 to $-$6.4, while models find C stars for
$M_{bol} \leq -4.7$ at solar metallicity, and for $M_{bol} \leq
-4.5$ at $Z = Z_{\odot}/3$. Moreover, in commenting Figs.
\ref{Fig1} and \ref{Fig3} we already noticed that the short
wavelength filters are insufficient to estimate the magnitudes of
red C-rich sources, enshrouded by dust. C-star bolometric
magnitudes below $M_{bol}$ = $-$5 do exist in LMC \citep[see
e.g.][]{vanloon97}, but they should not represent the {\it
average} population. In view of the fact that optical C(N)
sources always show C/O ratios close to 1, this can be easily
understood in terms of evolution: subsequent phases give birth to
redder objects, more C-rich (e.g. classes CV5$-$CV6 of BKR2),
radiating their flux mainly in mid-IR and having a luminosity
function shifted at higher luminosities (see e.g. VL2).

Compared to previous work on galactic C-rich giants by BKR1 \&
BKR2, our sample (with more homogeneous photometry and with a
wider coverage of IR wavelengths) further emphasizes the
importance of high luminosity C-stars. In the luminosity functions
(LFs) of the above authors (see e.g. Figure 5 of BKR1, where the
CV sources with the best distances were plotted) about 43\% of
CV3--CV6 sources are brighter than $M_{bol}$ = $-$5, while in our
sample (Fig. \ref{Fig8}) these account for about 60\% of the total
sample; similarly, 33\% of CV3--CV6 sources in the quoted papers
lay below $M_{bol}$ = $-$4, while in our data their fractional
number is about 2\%. On the whole, our distribution is displaced
to higher luminosities due to a higher weight of observations at
long wavelengths.

Our sample of galactic stars integrates previous IR studies on
dust-enshrouded C stars in LMC \citep[][VL1,VL2]{vanloon98}. These
works were forcedly restricted to a relatively small number of
sources: in this limit, their bright LF had little overlap with
the fainter one of optically-selected C stars (VL1, VL2).
However, other selection criteria lead to intermediate average
magnitudes \citep[around $M_{bol}$ = $-$5: see][]{whitelock}. Our
Fig. \ref{Fig8} now shows that the LF of galactic C stars is in
fact continuous, unique and quite wide. A comparison with the
bottom panel shows that our global distribution looks pretty much
as a superposition, with gaps filled, of LFs previously obtained
for different samples of C stars in LMC.

As mentioned previously, remaining uncertainties on bolometric
magnitudes for the reddest stars, for which we have sometimes an
incomplete coverage of the IR flux, would further populate the
high-luminosity tail of the histogram, making our conclusions even
stronger. The few points we have at low luminosity ($M_{bol} \geq
-4$) correspond to the HC classes of BKR2. We refer to those
papers for a detailed list of the possible physical mechanisms
explaining them.

\begin{figure}[t!]
   \centering
   {\includegraphics[width=\hsize]{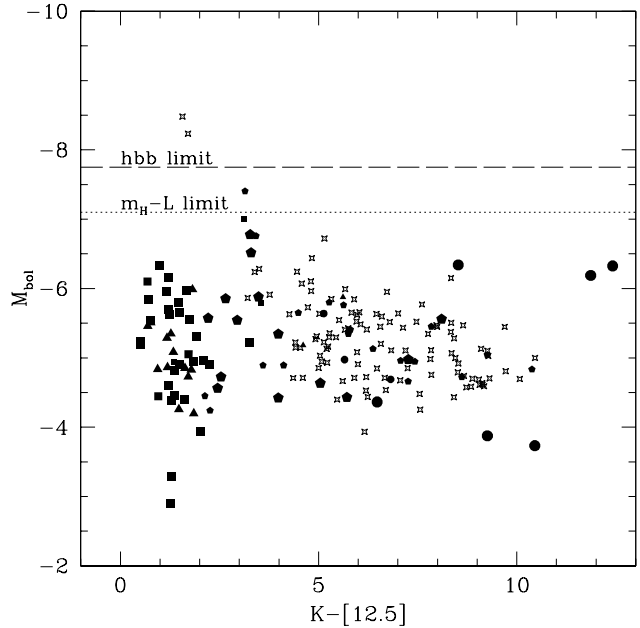}}
    \caption{The HR diagram of observed C stars adopting as a
    temperature indicator the near-to-mid IR colour K--[12.5], where
    contributions from both the photosphere and dust are present.}
\label{Fig9}%
    \end{figure}

\begin{figure}[t!]
   \centering
   {\includegraphics[width=\hsize]{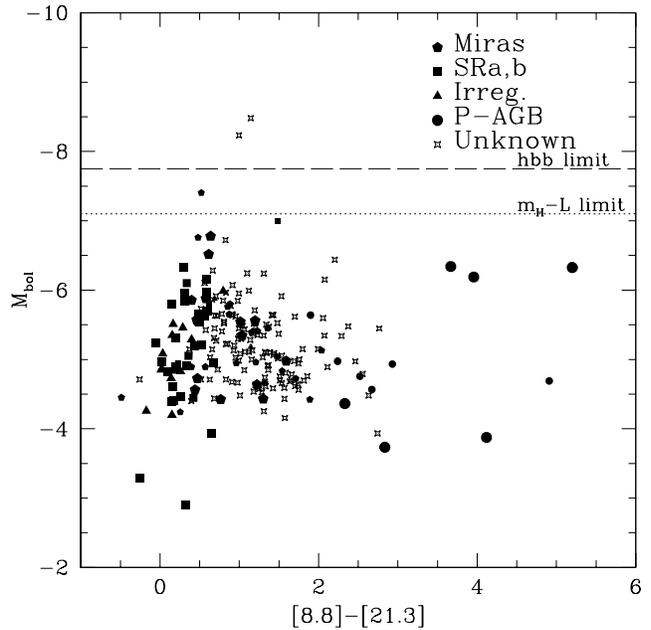}}
    \caption{The HR diagram of observed C stars in mid-IR, adopting the
    mid-IR colour [8.8]--[21.3] as abscissa.}
\label{Fig10}%
    \end{figure}

A few sources in Fig. \ref{Fig7} exceed model predictions for the
luminosity. Let us discuss this in some detail. We recall that a
well-defined relation exists on the AGB \citep{pacz, iben3,
wagenhuber} between the core mass ($m_{\rm H}$) and the luminosity
($L$). Only if $m_{\rm H}$ reaches the Chandrasekhar limit can an
AGB star attain the (slightly model-dependent) limit of
$-$7.1$^m$. Values higher than this require an extra source of
energy, as in the case of massive AGB stars (M $\ge$ 5 \ms), which
develop very high temperatures at the base of the convective
envelope, allowing the onset of hbb. In models by \citet{karakas},
AGB stars with hbb can in fact reach M$_{bol} \sim -$7.75. Normal
AGB stars (i.e. without hbb) must instead remain well below the
$m_{\rm H}-L$ limit, as $m_{\rm H}$ never reaches the
Chandrasekhar limit, due to efficient mass loss.

This is even more true for C stars: very bright Intermediate Mass
Stars (M = 5--9 \ms) can hardly become carbon rich, even in the
absence of hbb, as any $^{12}$C dredged-up from He-burning layers
is diluted over several solar masses of oxygen-rich material in
the convective envelope \citep[see
e.g.][]{busso99,gallino98,abia1,abia2}.

Recently, C stars more massive than 4--5 \msb have been obtained
by \citet{karakas}, but only for relatively low metallicities (Z =
0.004 to Z = 0.008). Such short-living stars should however be
extinct by now. Formation of very bright C stars is in principle
possible in the evolution of high-mass AGB objects (say, 6 to 9
\ms), when hbb ceases at the end, and if mass loss suitably limits
the envelope mass, so that the last few thermal pulses make the
photosphere become C rich \citep{frost8,karakas3,karakas}.
Statistically, however, these massive C stars should be extremely
rare, both for their low weight in the initial mass function and
for the very short duration of the theoretically permitted phases.
We can further notice that in Fig. \ref{Fig9} all the points
representing known AGB stars with reliable distance estimates
(large dots) stay (with one exception) below the limit for
moderately massive stars ($M \le 5 M_{\odot}$); points beyond this
level can therefore be simply explained as due to erroneous
classifications (C-rich supergiants, e.g. of Wolf-Rayet type).
Alternatively, they might be AGB stars whose luminosity has been
overestimated due to poor distances or to uncertainties induced
by variability.

HR-like diagrams can be constructed also using mid IR colours that
include the effects of dust. An example is in Fig. \ref{Fig9}: it
shows that there are no semiregulars or irregulars  for K$-$[12.5]
$\ge$ 3 and no Miras for K$-$[12.5] $\le$ 2; large IR excesses are
associated to Mira variability or to post-AGB stars, or to unknown
objects near the AGB end. Probably, the spread in magnitudes for
any colour is also an outcome of a spread in stellar masses.

An interesting piece of information is added by Fig. \ref{Fig10},
where the HR diagram is constructed using a purely mid-IR colour
as abscissa. Here we are looking only at dust, photospheres should
not be seen: we don't expect much further reddening on the
emission from dust, which re-radiates the stellar flux. Indeed,
the area covered by known long period variables is now quite
small, defining a narrow sequence for sources along the AGB, and a
limited colour extension of some more evolved sources, with a
maximum colour for Miras of [8.8]$-$[21.3] = 2. All known stars
redder than this are post-AGB objects; some unknown sources
observed by MSX are present in the reddest area, and we suggest
that they be classified as stars at the termination of the AGB
stage or above it, in evolution toward the ejection of a planetary
nebula. Large values of the mid-IR colours are in fact expected
when the relatively warm dust of AGB stars becomes cooler, forming
the extended (and often detached) shells of post-AGB stars.

\section{Mass Loss}

\begin{figure}[t!]
   \centering
   {\includegraphics[width=\hsize]{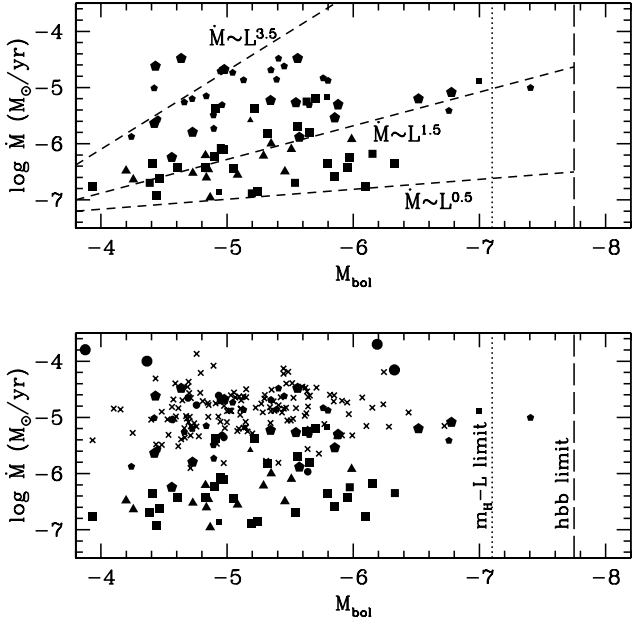}}
    \caption{Top panel: the bolometric magnitudes for AGB C-rich stars
of known variability type, as indicated, plotted as a function of
mass loss rates. Bottom panel:  same as before, but including
unknown and post-AGB sources.}
\label{Fig11}%
    \end{figure}

Consider now Fig. \ref{Fig11} (panel a); it shows the relation
between mass loss rates and bolometric magnitudes.  In panel a),
referring to irregulars, semiregulars and Miras, there is a
general increase of mass loss rates with luminosity, which cannot
however be reduced to a simple analytical formula (as examples,
power-law relationships with three representative exponents have
been added). The fact is that mass loss rates and luminosity
should be correlated as, while becoming more luminous, AGB stars
become also increasingly cool, their pulsation becomes stronger
and both facts should power more intense stellar winds. The spread
in panel a), partially confusing the relation, should then be
ascribed to the presence of other variables, possibly the mass,
and/or the fact that different mechanisms power stellar winds in
different evolutionary stages. This evolutionary hypothesis was
suggested by VL2 and \citet{vanloon05} as the cause of a spread
similar to the one we show; they in particular indicated that a
switch from single scattering \citep{jura4} to multiple
scattering \citep{gail86} of photons on dust grains as the star
evolves can explain mass loss rates in excess of the momentum of
radiation $L/c$. Notice that the points representing Miras stay
in the upper part of the diagram, showing a steeper dependence of
mass loss rates on luminosity. A sharp steepening of mass loss
rates, up to a final superwind phase, is in fact necessary to let
the whole envelope be ejected before the AGB termination.
Including post-AGB stars and unknown sources in the plot, as done
in Fig. \ref{Fig11}, panel b), introduces further complexities.
We believe we have here two problems. On one side, as already
mentioned, the reddest objects radiate part of their flux at
wavelengths longer than 45 $\mu$m, so that the luminosity we
estimate for them is only a lower limit. On the other side, these
sources are leaving (or have just left) the AGB and should have
huge circumstellar envelopes with possibly detached shells
escaping our spatial resolution. Any correlation is then hampered
by poor spatial and/or spectral coverage.

\begin{figure}[t!]
   \centering
   {\includegraphics[width=\hsize]{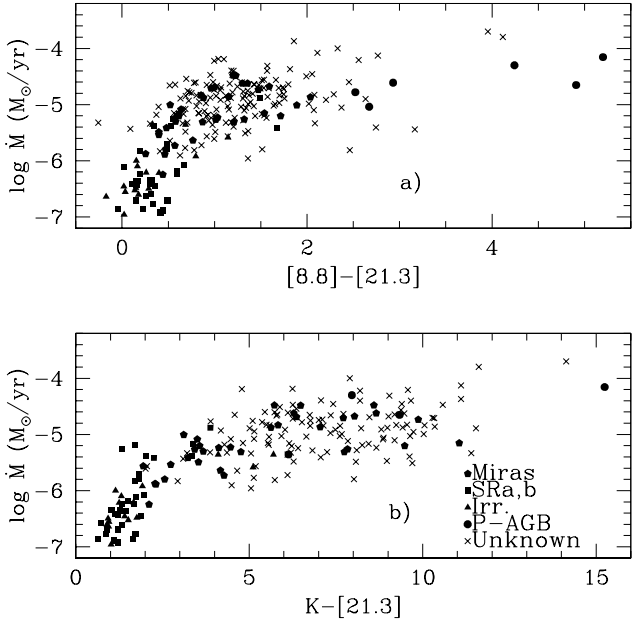}}
    \caption{Relations linking the mass loss rates of C-rich AGB and post-AGB
sources to the IR colours up to 21 $\mu$m.}
\label{Fig12}%
    \end{figure}

A good monitor of the mass loss rate is instead available from the
IR colours. This is illustrated in Fig. \ref{Fig12}. In
particular, when mid-IR colours are used, especially if including
the 21.3 $\mu$m (with the characteristic emission feature of
C-rich post-AGB stars), then the variation of mass loss rates
with the stellar classes (and with evolution, if the hypothesis
that Miras are on average quite evolved holds) is clearly
distinguishable (panel a). All known long period variables remain
at values of the [8.8]$-$[21.3] $\mu$m colour below 2, while
post-AGB objects display much redder colours, probably due to cold
dust formation. Again, transition objects exist among the sources
of unknown classification. A similar effect is seen from the
K$-$[21.3] colour (panel b), but here the separation of post-AGB
stars from the rest is less clear, and a rather continuous
distribution is seen. In general, the growth of the IR excess
linking Miras to post-AGBs seems to be rather smooth, with no
obvious break (e.g. for the superwind phase that is expected to
terminate the AGB evolution).

Data like those discussed in this paper might be useful to infer
which part of the mass return to the interstellar medium is due to
carbon stars. This would require to integrate the mass lost by AGB
stars over an Initial Mass Function and a Star Formation Rate for
low and intermediate mass stars and to estimate the fractional
contribution of carbon-rich phases. We shall therefore reconsider
this task in a subsequent issue of this analysis, after M giants
are included in the discussion.

\section{Conclusions}

In this paper we have presented the first issue of a series of
works on the IR properties of galactic AGB and post-AGB stars,
based on archived data from the 2MASS, MSX (6C Point Source
Catalogue) and ISO-SWS experiments. In particular, we analyzed a
sample of more than 230 C-rich AGB stars, for which estimates of
distances, mass loss rates and colours from 1 to 21 (and sometimes
45) $\mu$m could be collected. We found that uncertainties still
affect distances, except for a limited subset of objects with
astrometric-quality data. Many sources were found to be extremely
red, with SEDs extending to wavelengths unreachable by
ground-based observations: in particular, we showed that, starting
from the Mira stage, AGB carbon stars begin to radiate most of
their flux beyond 10 $\mu$m. This excludes the possibility of
deriving meaningful results on AGB luminosities by extrapolating
fluxes obtained at optical or near-IR wavelengths (a practice that
is unfortunately common even today). Using the best group of data,
for which reliable distances and the whole SEDs up to 45 $\mu$m
were available, we derived precise bolometric corrections, to be
applied to the other sources, and thus we constructed the
colour-magnitude diagrams in near and mid IR. The results suggest
that AGB carbon stars of the Galaxy are objects of moderate mass
(below about 5 \ms), and reach bolometric magnitudes between
$-$4.5 and $-$6.4. These findings confirm (at a higher metallicity
and with larger statistics) previous suggestions on Magellanic
Clouds by other authors. The luminosities derived are in good
agreement with those predicted by stellar models using the
Schwarzschild's criterion for convection and are at odd with the
belief (still persisting in some theoretical studies and
population-synthesis works), according to which observed C-star
luminosities would be lower than predicted by canonical stellar
evolution.

We also analyzed mass loss rates and their relation with the
magnitudes: we suggested that correlations between mass loss and
luminosity are hard to derive for post-AGB objects, probably due
to insufficient spatial and spectral coverage of their emission.
For AGB stars instead, correlations of mass loss rates with
luminosity exist, but reveal dependence by more than one
parameter. Here the stellar mass and the activation of different
mechanisms for powering mass loss should add their effects.
Clearer correlations emerged between the mass loss rates and the
IR colours, especially when the baseline was sufficiently extended
in wavelength. These correlations also provided tools for
classifying the evolutionary status of carbon stars.

\begin{acknowledgements}
We thank the referee, dr J.Th. van Loon, for an extensive and
helpful review, containing very relevant scientific advice. We
also thank N. Epchtein, O. Straniero and A. Chieffi for many
clarifying discussions. R.G. and M.B. acknowledge support in Italy
by MIUR, under contract PRIN2004-025729. The IRAIT project is an
Italo-French-Spanish collaboration funded in Europe by Polar and
Astronomical Agencies (in Italy by Programma Nazionale delle
Ricerche in Antartide and Istituto Nazionale di Astrofisica).

This research has made use of the SIMBAD database and the VizieR
service (CDS, Strasbourg, France), and the IRSA (NASA/IPAC
InfraRed Science Archive) database (USA). In particular archived
data from the experiments MSX, ISO-SWS and 2MASS were used.
$\bullet$ The processing of the science data of the Midcourse
Space Experiment (MSX) was funded by the US Ballistic Missile
Defense Organization with additional support from NASA Office of
Space Science. $\bullet$ The Infrared Space Observatory (ISO) is
an ESA project with instruments funded by ESA Member States
(especially the PI countries: France, Germany, the Netherlands and
the United Kingdom) and with the participation of ISAS and NASA.
$\bullet$ 2MASS (Two Micron All Sky Survey) is a joint project of
the Univ. of Massachusetts and the Infrared Processing and
Analysis Center (IPAC) at California Institute of Technology,
funded by NASA and the NSF (USA).
\end{acknowledgements}

\bibliographystyle{aa}

\Online
\begin{table*}[t!!]
            \centering
            \caption[]{Relevant data of our sample of carbon stars with
            astrometric, or reliable, distance estimates.
            Here and in next tables the following rules apply for
            labelling. i) For the variability type, M means Mira;
            S means Semiregular; I means Irregular (P is used for all
            Post-AGB stars, ''-'' for all stars of unknown
            variability). ii) References for mass loss are quoted in the
            following way:
            B: stands for \citet{bergeat05}; L is for \citet{loup}; G is for
            \citet{groenewegen02b}; Mei1998 is for \citet{meixner};
            Men2002 for \citet{menshchikov}; S1997 for \citet{skinner};
            Jam1991 for \citet{jaminet}.
            iii) References for distance are: B for \citet{bergeat05};
            L for \citet{loup}; G for \citet{groenewegen02b};
            S for \citet{schoier01}, using methods from \citet{groenewegen02b};
            S* for the same authors, when using methods
            from \citet{loup}; Mei1998 for \citet{meixner};
            Men2002 for \citet{menshchikov}; S1997 for \citet{skinner};
            Hon2003 for \citet{hony}; Bains2003 for \citet{bains}; Phill1998 for
            \citet{phillips}; K: method by \citet{kastner92}.}
            \label{tab1} %
            \vspace{-0.4cm}
            \includegraphics[width=\hsize,angle=0]{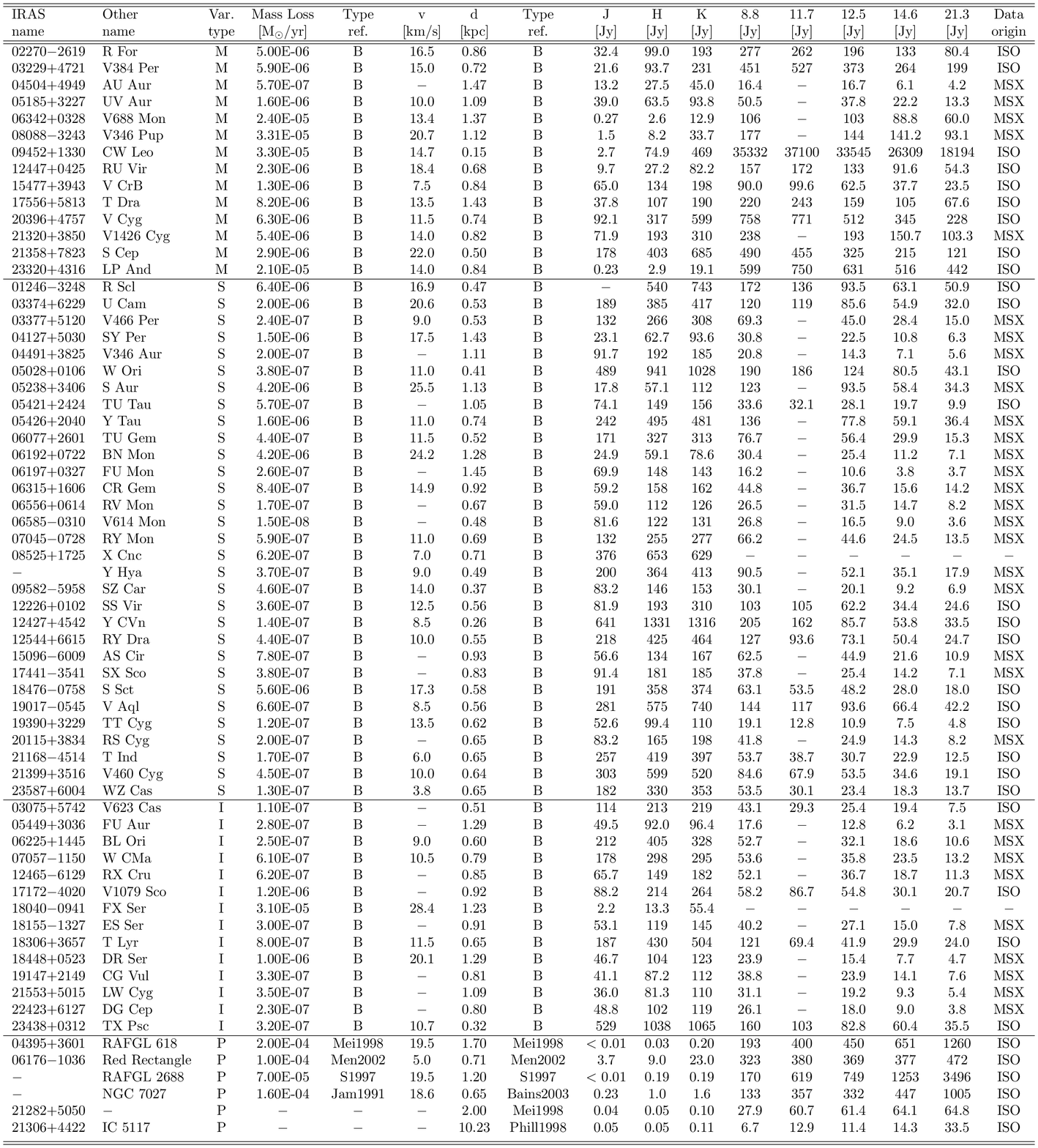}
\end{table*}

\begin{table*}[t!!]
            \centering
            \caption[]{Relevant data for AGB C stars of known variability type with
            non-astrometric distances.}
            \label{tab2} %
            \vspace{-0.4cm}
            \includegraphics[width=\hsize,angle=0]{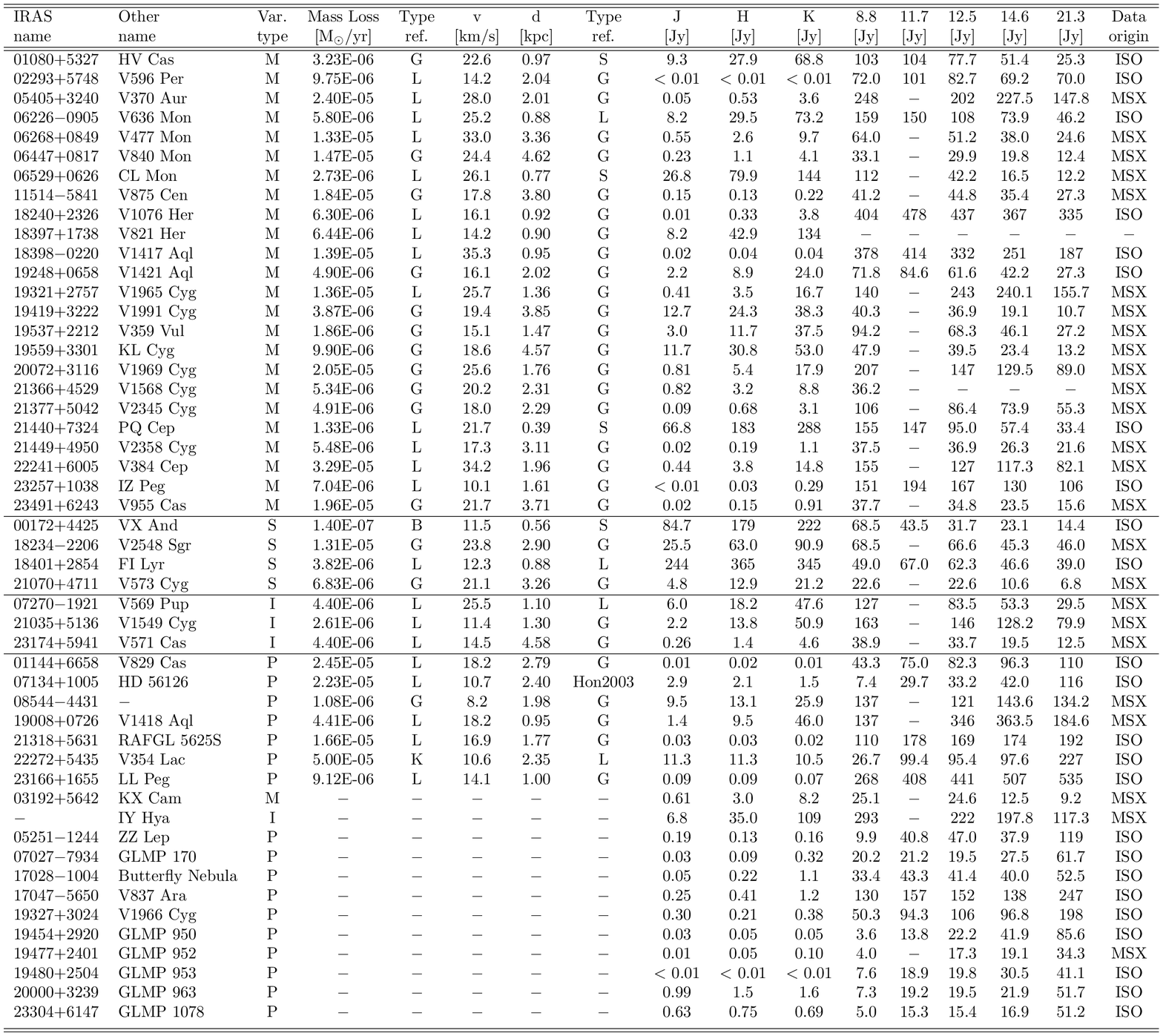}
\end{table*}

\begin{table*}[t!!]
            \centering
            \caption[]{Relevant data for AGB C stars of unknown variability type and with
            non-astrometric distances. First part.}
            \label{tab3} %
            \vspace{-0.4cm}
            \includegraphics[width=\hsize,angle=0]{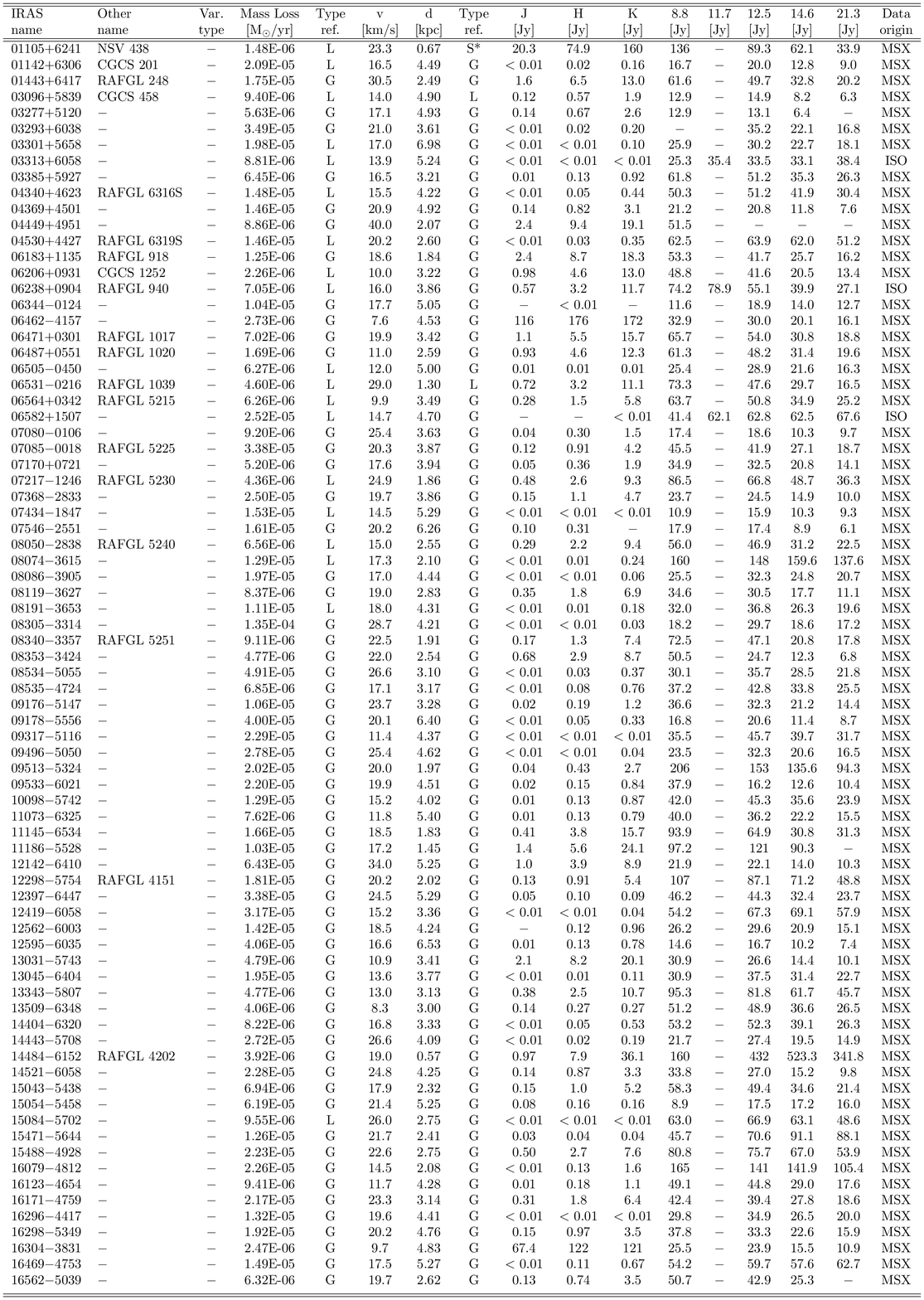}
\end{table*}

\begin{table*}[t!!]
            \centering
            \caption[]{Relevant data for AGB C stars of unknown variability type and with
            non-astrometric distances. Second part.}
            \label{tab4} %
            \vspace{-0.4cm}
            \includegraphics[width=\hsize,angle=0]{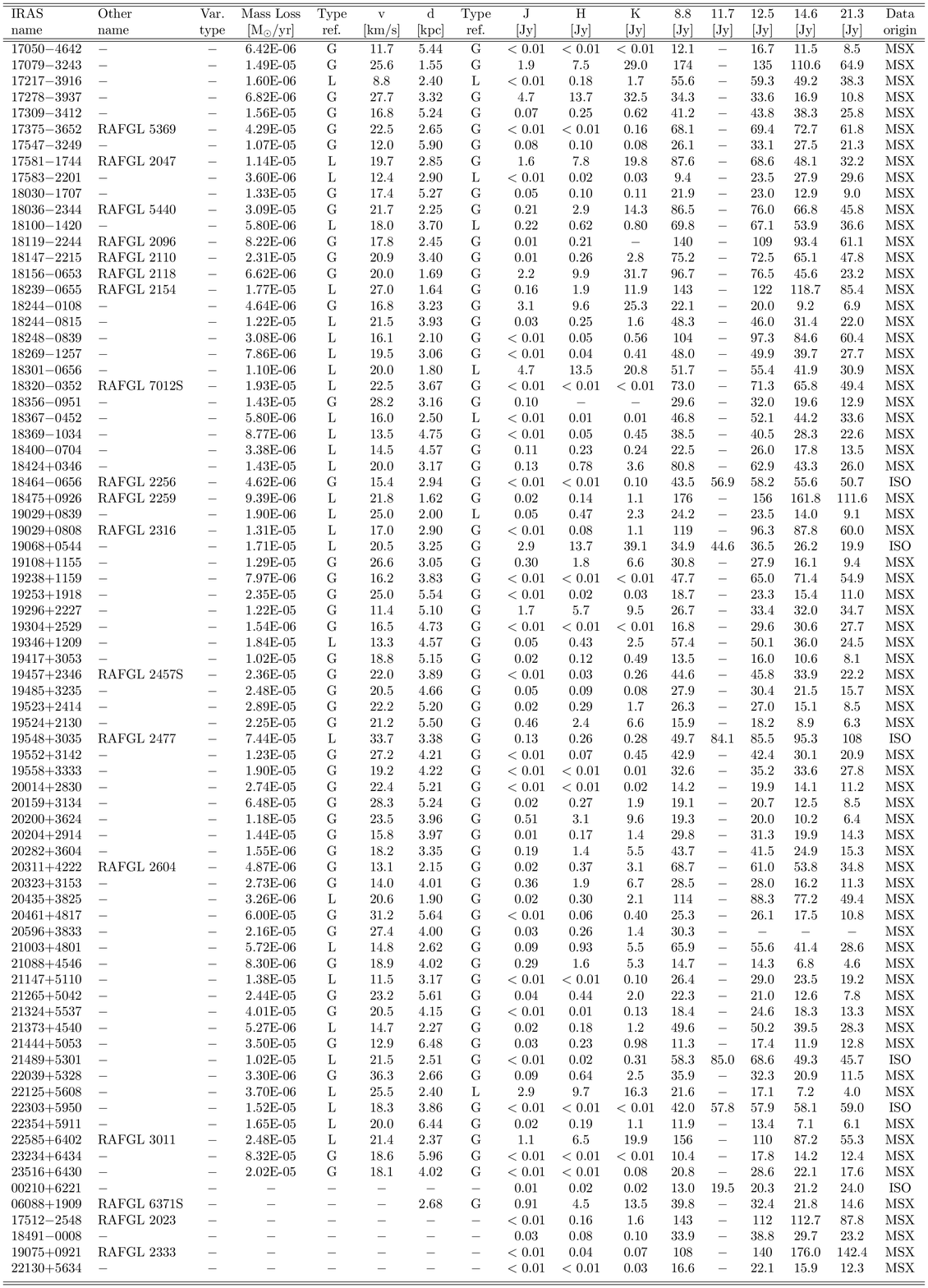}
\end{table*}

\end{document}